\documentclass{sig-alternate}
\usepackage{graphicx}
\usepackage{url}
\usepackage{subfigure}
\begin{document}
%
\title{Modelling page-view dynamics on Wikipedia}
\numberofauthors{4} 
%
\author{
\alignauthor Marijn ten Thij\\
       \affaddr{University of Twente}\\
       \affaddr{the Netherlands}\\
\alignauthor Yana Volkovich\\
       \affaddr{Barcelona Media Foundation}\\
       \affaddr{Spain}\\
\and
\alignauthor David Laniado \\
      \affaddr{Barcelona Media Foundation}\\
      \affaddr{Spain}\\
\alignauthor  Andreas Kaltenbrunner\\
       \affaddr{Barcelona Media Foundation}\\
      \affaddr{Spain}
      }
\maketitle

\begin{abstract}
  We introduce a model for predicting page-view dynamics of promoted
  content. The regularity of the content promotion process on
  Wikipedia provides excellent experimental conditions which favour
  detailed modelling. We show that the popularity of an article
  featured on Wikipedia's main page decays exponentially in time if
  the circadian cycles of the users are taken into account. Our model
  can be explained as the result of individual Poisson processes and
  is validated through empirical measurements. It provides a simpler
  explanation for the evolution of content popularity than previous
  studies.
\end{abstract}
\section{Introduction}\label{sec:introduction}
The social media boom gave a birth to a wide range of studies about
online traces generated by Internet users.  One of the important
research targets addressed by these studies is the analysis and
prediction of the dynamics of content popularity.  Historically, most
of these scientific works focused on the analysis of content
generated on blogging~\cite{duarte07,kumar2005bursty}, later
microblogging~\cite{lehmann2012dynamical},
video-sharing~\cite{Szabo2010} and news-sharing
platforms~\cite{kaltenbrunner_LAWEB2007}. However, in many cases the
studies reflect only the behaviour of registered users or focus on a
website of interest only for a specific community. Here we analyse
instead a website of general interest and address the problem of
understanding online usage and popularity patterns through a
large-scale analysis of the behaviour of the visitors of Wikipedia,
the sixth most visited
website\footnote{\url{www.alexa.com/siteinfo/wikipedia.org}}.

Wikipedia has an estimated number of 365 million monthly readers
worldwide~\cite{TED2010}. Although many studies analysed editing and
commenting activity on Wikipedia,
e.g.~\cite{kaltenbrunner2012WikiSym,ortega-wikipedia-2009,suh2009singularity,circadian},
there are not many quantitative works focusing on the Wikipedia usage
by the Internet users. A few studies explore Wikipedia views as an
information source in order to detect and predict events in real
world. Osborne et al.~\cite{osbornebieber} used a stream of Wikipedia
page-views to improve the quality of discovered events in Twitter, and
Mesty\'an et al.~\cite{Mestyan2012movie} predicted the popularity of a
movie by measuring the activity level of editors and viewers of the
corresponding Wikipedia entry. Finally, two studies
\cite{ratkiewicz2010a,reinoso2012wiki_traffic} analysed how the
Wikipedia traffic data is influenced by external and internal events.

The goal of this work is to examine the temporal evolution of the
popularity of promoted content on Wikipedia.
The competition for attention between numerous amounts and
different types of content makes it difficult to find sufficiently
regular and repetitive conditions which allow to predict how much
attention will be devoted from Internet users to a piece of content.
Some studies~\cite{Szabo2010,Wang:2012:UCO:2339530.2339573,Wu06112007}
tried to address this question by analysing media-sharing platforms
such as Digg or Youtube.  These platforms rank and categorise content
based on previous popularity and user votes, which leads to
rich-get-richer bias in the number of views and in the duration of the
promotion time.

The concept of content promotion on Wikipedia is
distinctively different. Similar to many online platforms, on
Wikipedia some of the articles get promoted to the \emph{Main
  page}\footnote{\url{http://en.wikipedia.org/wiki/Main_Page}}.
Promoted articles on Wikipedia are generated and managed through
online collaboration and a fixed number is shown to the online
audience during a fixed amount of time. This predefined exposure
duration together with the fact that only one article is promoted per
day makes the Wikipedia promotion mechanism unique in its regularity
compared to other large and popular social media platforms. It provides thus
excellent experimental conditions.

\begin{figure}[!t]
\centering
\includegraphics[width=\columnwidth]{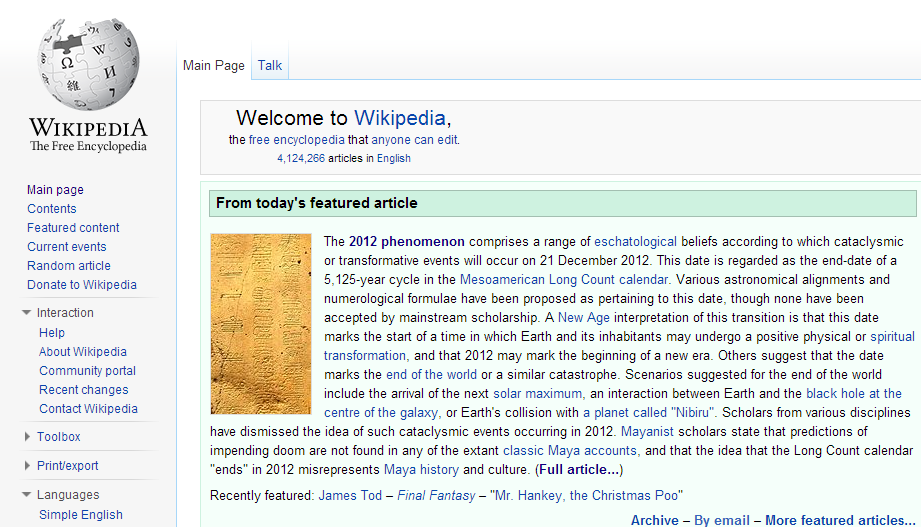}
\caption{The promoted article on the English Wikipedia's \emph{Main
    page} of December 20, 2012.\label{fig:MP}}
\end{figure}

Figure~\ref{fig:MP} presents an example of the \emph{Main page} of the
English Wikipedia. Every Wikipedia user (although preference is given
to the articles' primary editors) can nominate articles with certain
characteristics (\emph{featured articles}) to the pool of possible
future promoted articles on a specific
page\footnote{\url{http://en.wikipedia.org/wiki/Wikipedia:Today's_featured_article/requests}}.
We refer to the article placed under the headline ``From today's
featured
article''\footnote{\url{http://en.wikipedia.org/wiki/Wikipedia:Today's_featured_article}}
as \emph{promoted}. Every day, at 00h(UTC), a new promoted article is placed
on the \emph{Main page} together with links to the three articles
promoted during the previous three days (see ``Recently featured'' at
the bottom left of Figure~\ref{fig:MP}).  The ``today's promoted''
articles are also sent by e-mail to subscribers.





\section{Dataset}\label{sec:dataset}
We retrieved the page-view values from a database provided by
\emph{Wikimedia}\footnote{\url{http://dumps.wikimedia.org/other/pagecounts-raw/}}.
This database contains one file per hour, listing the total
number of views to a page during that hour, provided it received at
least one view.  We extracted the page-view data between December 9,
2007 at 18h(UTC) and March 31, 2010 23h(UTC), for a total of 844
days. Note that this database is not entirely complete: for some hours
there is missing data or there is more than one entry for a single
article. For the latter we just sum these entries. Finally, we assume
that views to articles which come through redirects are also
registered in the data of the target page.

For comparison with the general access pattern to Wikipedia we also
used the Wikipedia dump from March 12, 2010 and extracted the view
data of all Wikipedia articles which received at least one comment
until this date in their discussion page. This led to 871~395 articles
which accumulated $32\cdot 10^9$ views in total, i.e. on average
$38\cdot 10^6$ per day.


\section{Page-view Statistics}\label{sec:pagestats}
We describe here the hourly and daily number of page-views
on the English Wikipedia
and analyse then the popularity of a promoted article during the
promotion period.

\subsection{Circadian and weekly patterns}\label{subsec:circadianpatterns}
\begin{figure*}[!t]
  \centering
\includegraphics[width=0.32\textwidth]{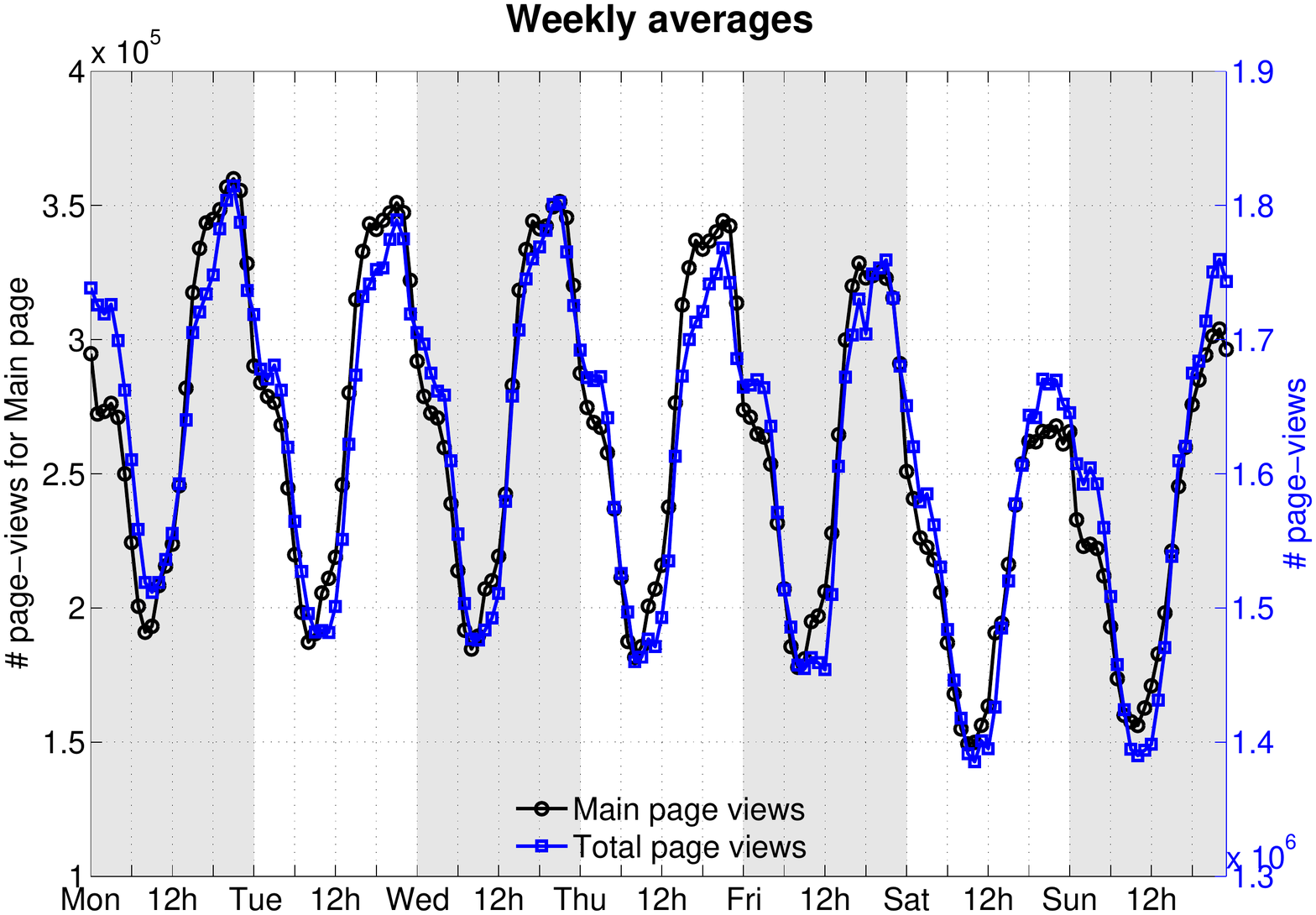}
\includegraphics[width=0.32\textwidth]{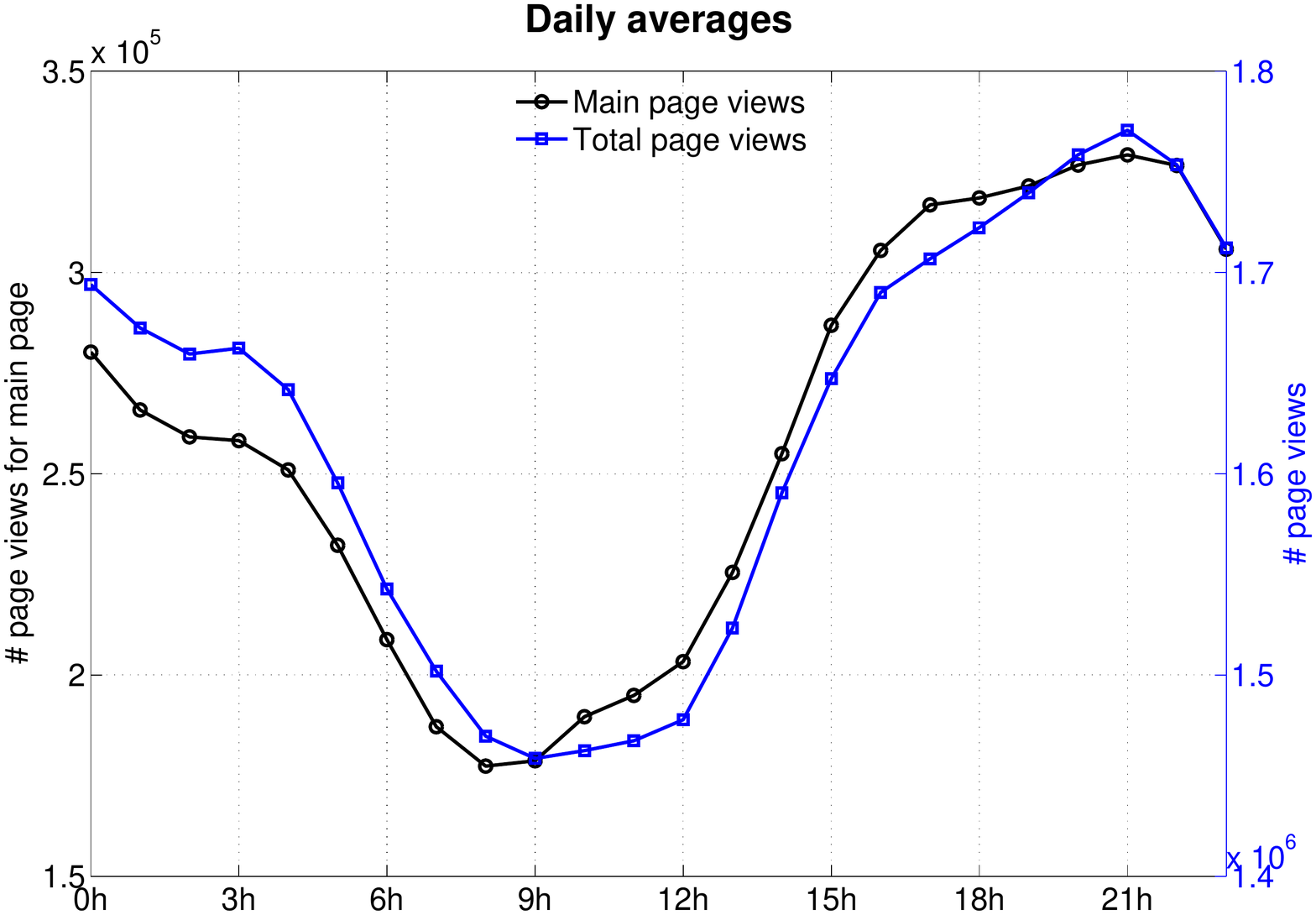}
\includegraphics[width=0.32\textwidth]{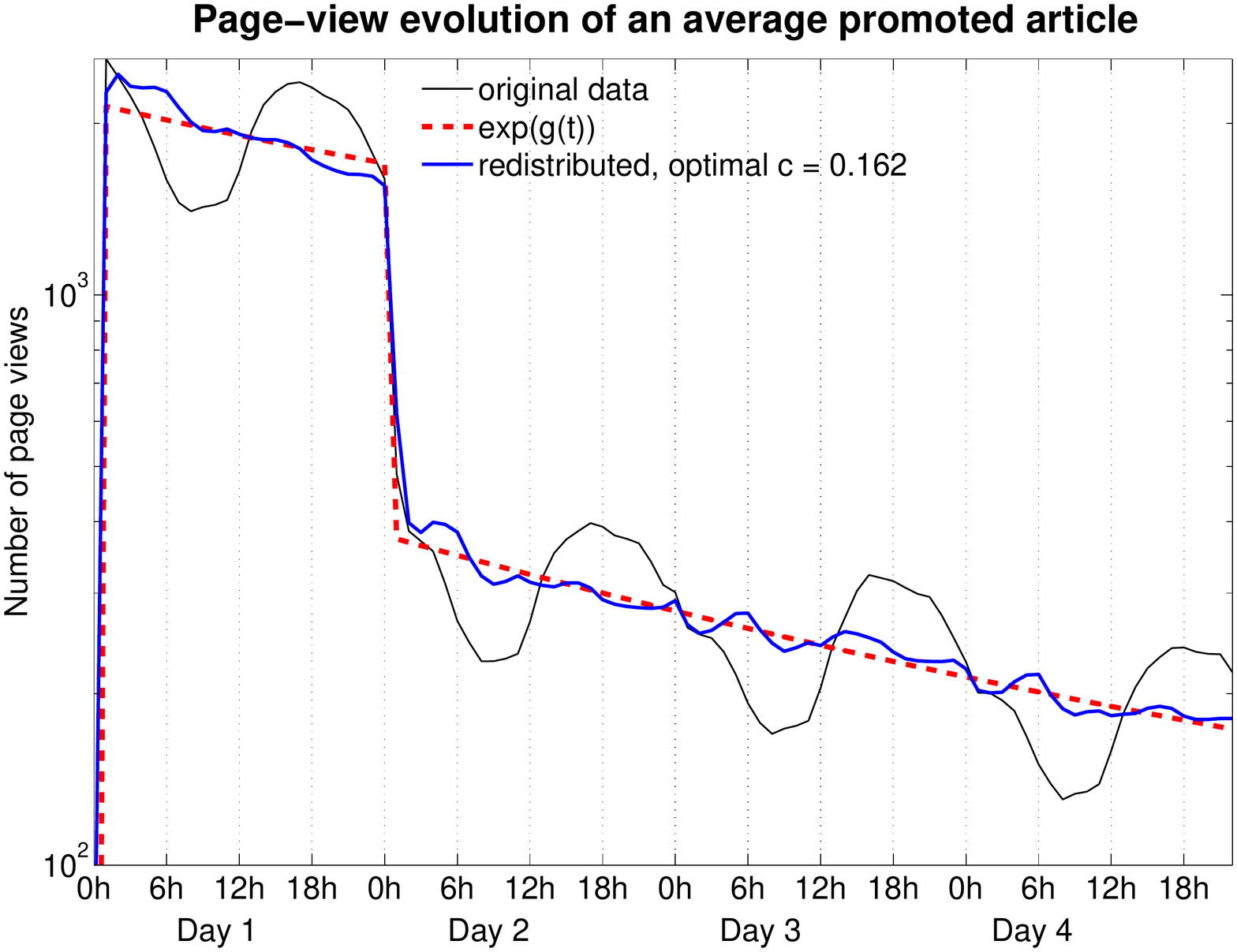}
\caption{Temporal patterns of the Wikipedia views per hour in
  total and for its \emph{Main page} \label{fig:view_patterns} (left, centre) and the
  average number of views of a promoted article arranged by different
  time scales \label{fig:redistributed} (right).}
\end{figure*}

In Figure~\ref{fig:view_patterns} we depict the number of visits per
hour to the English Wikipedia in general and also to its \emph{Main
  page} averaged by weeks (left) and by days (middle sub-figure). The
visits of the English Wikipedia vary between $1.4\cdot 10^6$ and
$1.8\cdot 10^6$ with an average of $1.6\cdot 10^6$ page-views per hour. For
the \emph{Main page} popularity we find on average $2.5\cdot 10^5$
visitors per hour.
The \emph{Main page} pattern is similar to the overall weekly
pattern. Both slightly decrease during weekends.

Figure~\ref{fig:view_patterns} (middle sub-figure) depicts the
circadian patterns of the Wikipedia page-views. We observe the lowest
activity between 07h and 10h(UTC) corresponding to the night hours in
the US. Similar circadian patterns but
for editing activity on the English Wikipedia were observed
in~\cite{circadian}.

\subsection{Promoted articles}
In the previous section we have analysed the temporal views of
Wikipedia articles and the Wikipedia's main page in general. In the
rest of this work we will focus on the page-view data for the promoted
articles only. Recall that in Wikipedia an article gets promoted for a
predefined period of 1+3 days (96 hours), which we call the exposure duration in
analogy to \cite{Wang:2012:UCO:2339530.2339573}. We restrict our
analysis and predictions only to these article exposure durations.

We select all the articles promoted in the time-span from January 1,
2008 to March 31, 2010, which have complete page-view data, i.e.
we obtained the number of views for every hour in their exposure
duration.
We omit the articles of \emph{Barack Obama} and \emph{John
  McCain},\footnote{This is the only occasion where 2 articles are
  promoted at once for the reason of the US presidential elections.}
promoted both\ on November 4, 2008. They show completely different
dynamics compared to the average article and would influence some of
the results reported below as they have the largest number of views
during the second day of exposure (once the presidential elections
were decided).
Thus, in total we use 684 promoted articles in this study.


By popularity of a promoted article we mean the number of views this
article receives during the exposure duration.  The right sub-figure of
Figure~\ref{fig:redistributed} depicts the average number of views
$v_t$ a promoted article attracts during the $t=1,\ldots,96$ hours of
exposure. The exposure period of a promoted article in Wikipedia can
be divided into four stages. At the first stage, during the first hour
after a page gets promoted, we witness a huge increase in the
article's popularity. This value $v_1$ is the largest for the average
promoted article. The second stage contains the remaining hours of the
first day of the promotion.  The third stage is characterised by the
sharp decay occurring after the original article gets replaced by the
new one. Finally, the last stage contains the view dynamics during the
3 days of being promoted in ``Recently featured''. Using this
stage-representation, we construct $g(t)$ (dashed red line in
Figure~\ref{fig:redistributed}, right) as a piecewise-linear approximation of
$\log(v_t)$.


\subsection{Circadian patterns correction}

Comparing the approximation $g(t)$ and the promoted article popularity
$v_t$ we notice that the main differences are caused by the circadian
patterns of Wikipedia views. To remove these variations we use a new
time scale in which every hour is measured in the number of views
rather than in minutes. This approach was introduced
in~\cite{Szabo2010} for the popularity of Digg stories. We
modify the original idea by removing a constant fraction $c$ of the
traffic data to emphasise the circadian patterns even more.

Formally, we denote as $m(t)$ the average number of \emph{Main page}
views for a given hour $t=1,\dots,24$. We define a new redistribution
parameter $T^*$ as follows:
\[T^*=\sum_{t=1}^{24}m^*(t)=\sum_{t=1}^{24}[m(t)- c\min_t m(t)],\]
where $c=\mbox{arg}\min \left[ \sum_{t^*=0}^{95}
  \left(\log\left(v_{t^*}\right) - g(t^*)\right)^2 \right]$ and
$v_{t^*}$ (depends on $c$) is the number of views of an average
promoted page at time $t^*$ in the new time scale defined by $T^*$. We
find that $c=0.162$ is the optimal value for ``decycling'' based on
the \emph{Main page} views. In other words one needs to remove
approximately 16\% of the minimum of the hourly traffic of the Wikipedia
\emph{Main page} to make an optimal correction of the circadian
patterns. A new hour $t^*$ is therefore the time interval which takes
the \emph{Main page} to accumulate from $\frac{t^*-1}{24}T^*$ to
$\frac{t^*}{24}T^*$.

In the rest of the paper we refer to the new time scale $t^*$ as
\emph{redistributed time scale}.  The blue line in
Figure~\ref{fig:redistributed} (right) depicts the average number of views
$v_{t^*}$ of the promoted articles in the redistributed time scale.
We observe the log-linear decreasing trend of the promoted article.
popularity.

\pagebreak
\section{Model}\label{sec:model}
Based on the average view behaviour, i.e. the average number of views
per hour over all promoted articles, we propose a model which
describes the traffic dynamics of a selected promoted article during
its exposure duration. This model is defined by two parameters: a
constant interest-decay factor for all days of the promotion and
negative jump of the popularity after the first day of
exposure. The number of views $v_1$ a selected article receives during
the first hour of the promotion is used as the only input value of the
model. 

\subsection{Model definition}
The definition of the model is inspired by the shape of the page-view
behaviour pattern, or more exactly by the normalised number of views
per time unit $w_{t^*}=v_{t^*}/v_{1^*}$ in rescaled time (with the
circadian cycle removed).\footnote{We use $v_t$ or $w_t$ when
  referring to data and $\hat{v}_t$ or $\hat{w}_t$ for the model
  curves. Recall also that the subindex $t$ stands for the real time
  and $t^*$ for the redistributed time.}

Based on the log-linear fit of $w_{t^*}$ and using $w_{1^*}=1$ we
define the model as:
\begin{eqnarray*}
\hat{w}_{1^*}&=&1 \\
\hat{w}_{t^*}&=&\beta_{t^*}\cdot\hat{w}_{t^*-1} \mbox{ for } t^*=2,\dots,95
\end{eqnarray*}


Previously, we have described the four stages of the exposure life of
a promoted article on Wikipedia. Using these definitions we set a
temporal factor $\beta_{t^*}$ as $\beta_{t^*}={\gamma}$ for $t^*=25$
and $\beta_{t^*}={\beta}$ for other $t^*$'s, i.e. for $2\leq t^* \leq
24$ and $ 26\leq t^* \leq 95$. The constant factor $\beta$ models the
decay of the number of page-views in a typical hour of the exposure
duration, i.e. while the article is promoted on the \emph{Main page}. The
factor $\gamma$ states for the negative jump in the number of views
after the promoted article gets moved to ``Recently featured''
position. Thus, we model the shape of the article popularity by stage:
the first stage of the promoted article is characterised by
$\hat{w}_{1^*}$, the second by the interest-decay factor $\beta$, the
third by $\gamma$, and the fourth again by the same factor $\beta$. To
summarise, we model the normalised number of views of the promoted
article during the $t^*$-th redistributed hour as follows:
\[
\hat{w}_{t^*}=\left\{\begin{array}{rcr}
    {\beta^{t^*-1}} & \mbox{for} &2\leq t^* \leq 24;\\
    {\gamma\beta^{t^*-2}} & \mbox{for}& 25 \leq t^* \leq
    95.\end{array} \right.
\]

\begin{figure}[!t]
\centering
\includegraphics[width=\columnwidth]{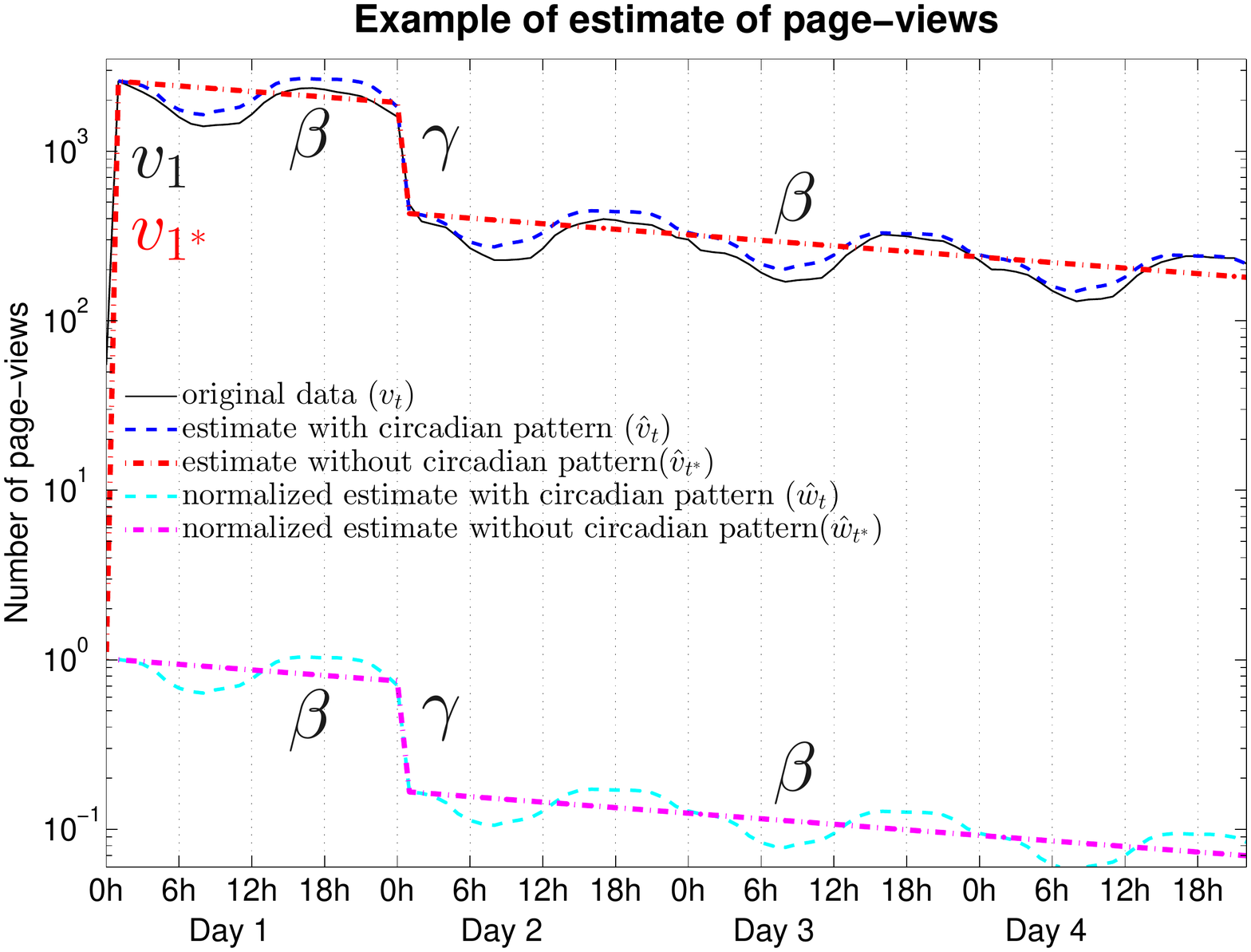}
\caption{Visualisation of the popularity model for promoted
  articles.\label{fig:prediction}}
\end{figure}

Finally, we use the reverse time-redistribution to find $\hat{w}_t,$
i.e.  the corresponding number for $\hat{w}_{t^*}$ but in the original
time scale.  We define the number of views of the promoted article
during $t$-th hour as
\[\hat{v}_t= v_{1^*}\cdot\hat{w}_t,\]
where $v_1$ and $v_{1^*}=v_1/\hat{w}_1$ are the numbers of views of
the promoted article after the first hour of exposure in the original
and redistributed time scales.

In Figure~\ref{fig:prediction} we draw a visual explanation of the
model. The dashed-dotted lines in magenta correspond to the model
curve $\hat{w}_{t^*}$ and the cyan dashed curve to its transformation
$\hat{w}_t$ into the real timescale. The multiplication of $\hat{w}_t$
with $v_1$ leads to the model approximation $\hat{v}_{t}$ (blue dashed
curve) of the original data $v_{t}$ (black
curve). Figure~\ref{fig:prediction} also depicts the rescaled model
$v_{t*}$ in the redistributed time scale as dashed dotted line in red.

\begin{figure}[!t]
\centering
\includegraphics[width=\columnwidth]{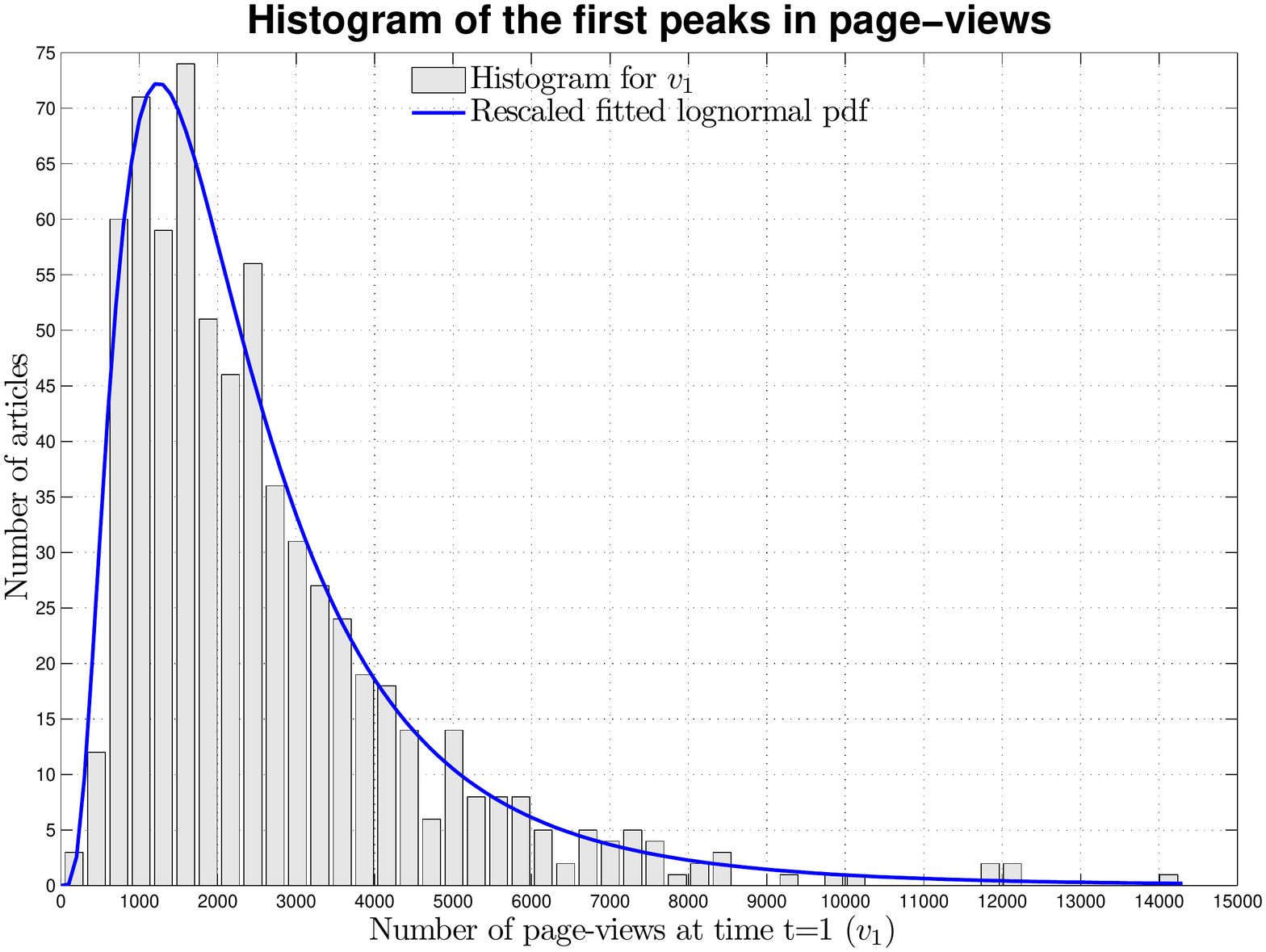}
\caption{Histogram of occurring values for
  $v_1$.\label{fig:hist_peaks}}
\end{figure}

The introduced model uses only the number of views during the first
hour of the exposure period $v_1$ as an input parameter.  In
Figure~\ref{fig:hist_peaks} we plot the histogram for $v_1$'s in our
dataset together with the log-normal fit ($\mu = 7.63$ and $\sigma =
0.71$). We have also investigated whether the values of $v_1$
correlate with the page-views of the corresponding articles before
being promoted. No such correlations were found.

\subsection{Model interpretation}
The model can be explained as the consequence of individual Poisson
processes at the user level. We assume that the users visit a promoted
article only once and thus, given that in a Poisson process the probability
of the first arrival time $T_1$ being larger than $t$ is
$$
\mbox{P}(t<T_1)=\exp(-\lambda t)=\beta^t,
$$
we can find that
$$
\mbox{P}(t-1<T_1\leq t)=\mbox{P}(t-1<T_1)-\mbox{P}(t<T_1)=\beta^{t-1}(1-\beta).
$$
The above formula corresponds to the likelihood that an individual
user visits a promoted article during the $t$-th hour of exposure on
the \emph{Main page}.  It is, apart from the constant factor
$(1-\beta)$, identical to the decay factor $\beta^{t-1}$ of our
model. This constant factor can be neglected as part of a normalising
constant. The parameter $\gamma$ corresponds to the decrease in the
likelihood of visiting an article after it has passed to the
``Recently featured'' section.

The actual number of page-views is determined by how many users get
curious about an article and visit it after they observe a link to it on
the \emph{Main page} (or receive it via an e-mail subscription). We
model the time distribution of these page-views, which is governed by
individual Poisson processes with rate $\lambda=-\ln(\beta)$.


\subsection{Parameter Estimation} \label{subsec:par_est} We estimate
the model's parameters $\beta$ and $\gamma$ by using page-view data of
the promoted articles on the English Wikipedia. To investigate the
stability of the parameter estimation we apply the estimation
algorithms for two sets of promoted articles. The first set $S_a$
contains all 684 articles and the second set $S_1$ the first 100
promoted articles by date of promotion.  We use $S_a$ to describe the
general view dynamics for promoted content on Wikipedia and $S_1$ to
predict the popularity of the 584 articles promoted
afterwards. 

We denote as $\hat{v}_{t^*}(\beta,\gamma)$ the predicted number of
views for given values of $\beta$ and $\gamma$ at redistributed time
$t^*$, and as $v_{t^*}$ the actual number of page-views at time $t^*$
for some promoted article $s\in S,$ where $S$ is either $S_a$ or
$S_1$. Then, we calculate parameters $\beta$ and $\gamma$ that
minimise the error:
\begin{equation}
  \left\{\beta,\gamma\right\}=\mbox{arg}\min\sum_{s\in S}\left[ \sum_{t^*=1}^{95}  \left[\log(\hat{v}_{t^*}(\beta,\gamma)) - \log(v_{t^*})\right]^2 \right]
\label{eq:min_obj}
\end{equation}
This yields to $\beta = 0.9874$ and $\gamma=0.2319$ for $S_as$ and to
$\beta=0.9877$ and $\gamma=0.2618$ for $S_1$. We observe that the
value of $\beta$ is very similar for the two set while there is a small
difference for $\gamma$.

\begin{figure}[!tb]
\centering
\includegraphics[width=\columnwidth]{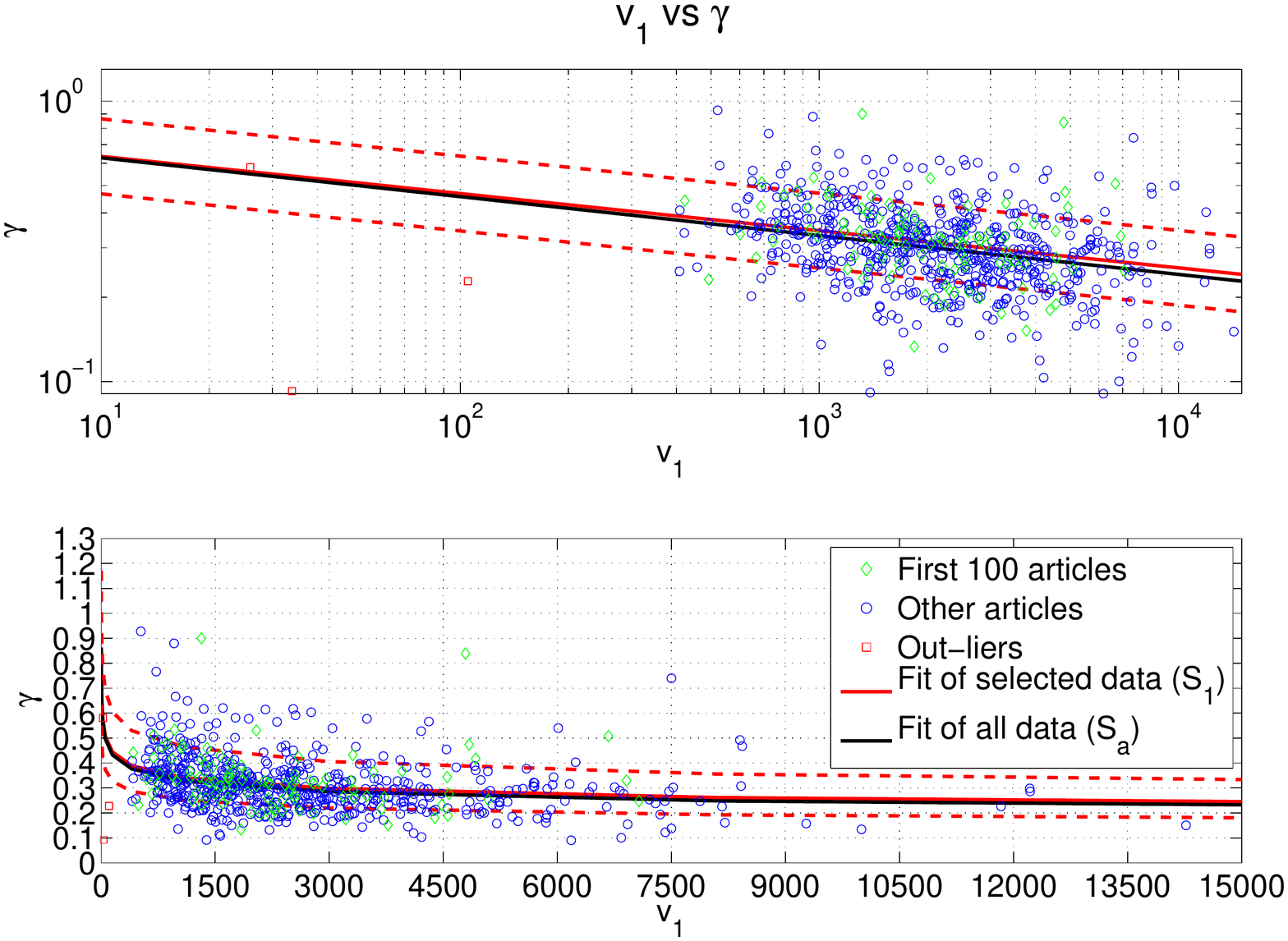}
\caption{Relations between $\gamma$ and $v_1$. The dashed red lines
  indicate the interval
  $[h(v_1)-\sigma,h(v_1)+\sigma]$.\label{fig:gamma_peak_100}}
\end{figure}

To investigate this further we drop the assumption that $\gamma$ can
be modelled as a constant factor for all articles and look at the
$\gamma$ parameters at the individual article level. Since $\gamma$
encodes the negative jump in the decay of user interest after the
first day of the exposure, we suggest that it should be correlated
with the overall popularity of the promoted article. To this end, we
compared the values of $\gamma$ with the total number of views a
promoted article receives during one day before the promotion date but
found no correlation between them.  Instead, we propose to define
$\gamma$ as function of the initial popularity $v_1$. We first find
that $\log(v_1)$ and $\log(\gamma)$ are negatively correlated
(Pearson's correlation coefficient is $-0.29$ for $S_a$ and $-0.25$ for
$S_1$), which is also indicated in
Figure~\ref{fig:gamma_peak_100}. Then, we derive a log-linear function
for $\gamma$: \[\gamma = C\cdot v_1^m,\] based on the observations
$\{\log(v_1),\log(\gamma)\}$ for the articles from set $S$, where $S$
is again either $S_a$ or $S_1$.  We rewrite the last equation in the
following form:
\begin{equation}
  \log(\gamma)= h(v_1)= m \cdot\log(v_1)+\log(C).
\label{eq:gamma_func}\end{equation}

Using the set $S_a$ we obtain $m=-0.138$ and $C=0.863$ for all
articles. We note that for estimation of parameters $m$ and $C$ we
omit the outliers\footnote{These are the articles \emph{Borobudur},
  \emph{Princess Beatrice of the United Kingdom}, \emph{Local
    Government Commission for England (1992)}, \emph{West Indian
    cricket team in England in 1988} and \emph{Attachment theory}.}
indicated as red squares in Figure~\ref{fig:gamma_peak_100}.  We also
perform the fitting for (\ref{eq:gamma_func}) on $S_1$ and obtain
$m=-0.132$ and $C=0.862$. Note that, although the initial estimates for
$\gamma$ were slightly different for $S_a$ and $S_1$, the parameters of
$h(v_1)$ are not. This can be also observed in the nearly overlapping
linear fits in Figure~\ref{fig:gamma_peak_100}. The reason for this
greater stability is that we now focus only on the size of the drop
and not on the effect of the choice of $\gamma$ on the minimisation of
the model error in the subsequent days as Equation~(\ref{eq:min_obj}) would do.

Comparing the estimated values for $\log(\gamma)$ with $h(v_1)$,
we find that
$\log(\gamma)\sim\mathcal{N}(h(v_1),\sigma^2)$. Therefore, we can
derive an interval in which the decay factor would lie with a given
probability. We use $[h(v_1)-\sigma,h(v_1)+\sigma]$ as this interval,
indicated by the dashed lines in Figure~\ref{fig:gamma_peak_100}.

Back to the model, we can now calculate $\hat{w}_{t^*}$ at time
$t^*$ as follows:
\begin{equation}
  \hat{w}_{t^*}=\left\{\begin{array}{rcr}
      {\beta^{t^*-1}} & \mbox{for} & 2\leq t^* \leq 24;\\
      { C\cdot v_1^m\cdot\beta^{t^*-2}}  & \mbox{for} & 25 \leq t^*
      \leq 95;\end{array} \right.
\label{eq:pred_v1}
\end{equation}
and then use the reverse time-redistribution to find $\hat{w}_t$.
Using $\hat{w}_t$ and
\begin{equation} \hat{v}_t= \frac{v_1}{\hat{w}_1}\cdot\hat{w}_t
\label{eq:pred_v1_2}
\end{equation}
we can obtain the estimated hourly progression $\hat{v}_t$ of the
page-views for $t=2,\ldots,95$.

\section{Popularity Prediction}\label{subsec:prediction}
As explained in the previous section, we use the first 100 promoted
articles (ordered by date of exposure) to learn the model
parameters and predict the popularity of the 584 Wikipedia articles of
our dataset promoted a posteriori. Thus, for each of these articles we
take the article's popularity after the first hour $v_1$ and use Eq.
(\ref{eq:pred_v1}) and~(\ref{eq:pred_v1_2}) with the
parameters $\beta=0.9877$ and $\gamma$ ($m=-0.132$, $C=0.862$) of $S_1$.

\begin{figure}[!tb]
\centering
\includegraphics[width=\columnwidth]{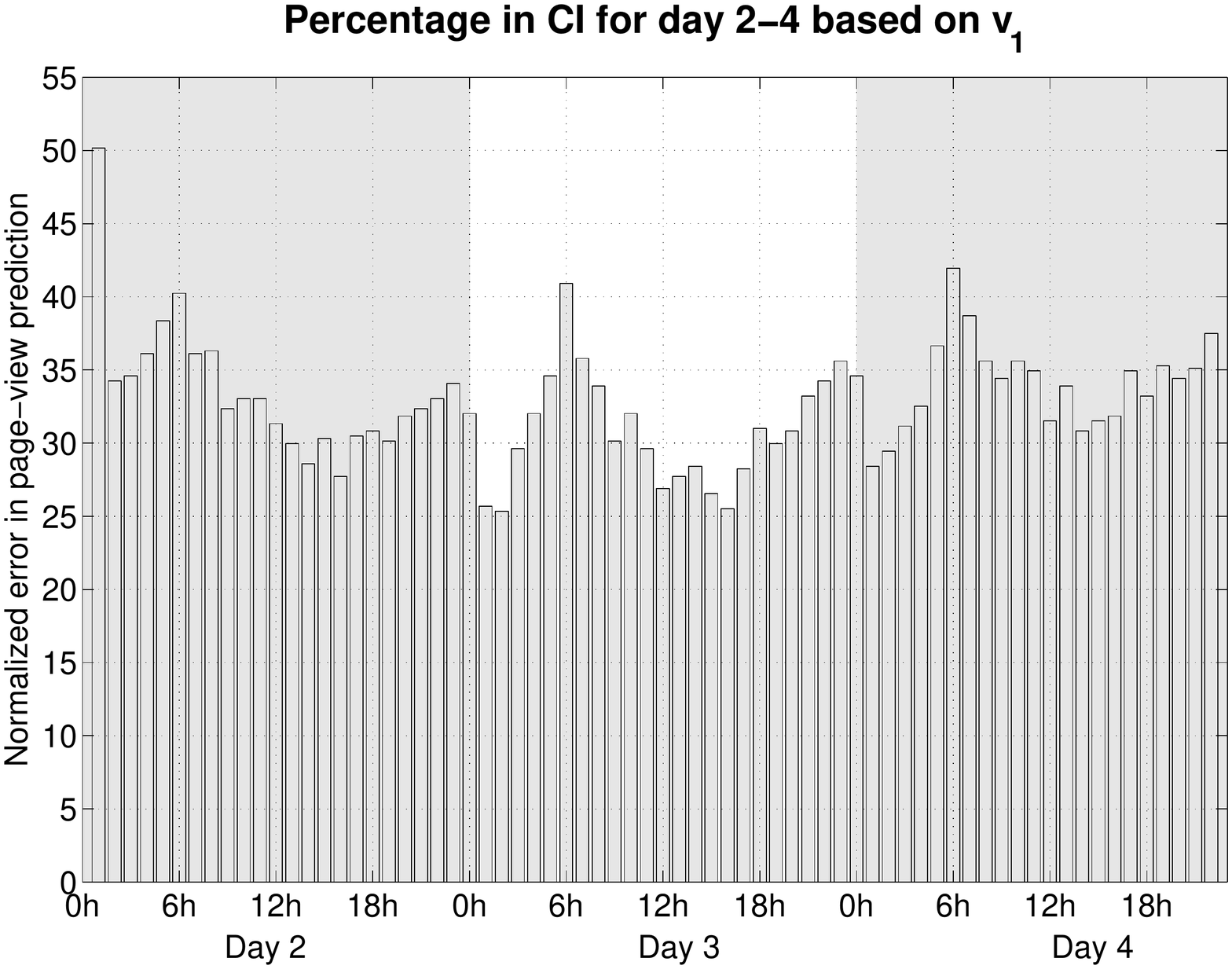}
\caption{Percentage of values $v_t$ in predicted interval for
  $t=25,\ldots,95$.\label{fig:perc_CI}}
\end{figure}

As we will discuss below for most of the promoted articles we are able
to obtain a good prediction for the page-view dynamics during the
first day of exposure. However, for the remaining days the number of
actual page-views $v_t$ does not always lie within the predicted
interval $[h(v_1)-\sigma,h(v_1)+\sigma]$ for $t=25,\ldots,95,$ as we
see in Figure~\ref{fig:perc_CI}. Thus, although in the 25-th hour we
correctly predict the page popularity for 50\% of the articles, in
general we observe a lower percentage of correct predictions. This is
caused by underestimating the decline of interest (or an
overestimation of $\gamma$) by our model and can be improved by
introducing the input parameter $v_{25}$, i.e. the value of the
promoted page popularity right after it is moved to the ``Recently
featured'' section, into our model.

Adjusting the prediction during the first hour of the second day of
the promotion with $v_{25}$ leads us to the following description of
the model:
$$
\hat{w}_{t^*}=\left\{\begin{array}{lcr}
    {\beta^{t^*-1}} & \mbox{for} & 1\leq t^* \leq 24;\\
    {\beta^{t^*-25}} & \mbox{for} & 25 \leq t^* \leq 95.\end{array}
\right.
$$
Here we use again the reverse time-redistribution to find $\hat{w}_t$
to obtain the predicted hourly page-views progression $\hat{v}_t$, for
$t=2,\ldots,95$ by calculating
\begin{equation}
  \hat{v}_t =\left\{\begin{array}{rcr} \frac{v_1}{\hat{w}_1}\cdot\hat{w}_t & \mbox{for} & 1\leq t \leq 24;\\
\frac{v_{25}}{\hat{w}_{25}}\cdot\hat{w}_t & \mbox{for} & 25 \leq t \leq 95.\end{array}\right.\nonumber
\end{equation}

\begin{figure}[!tb]
  \centering
\includegraphics[width=\columnwidth]{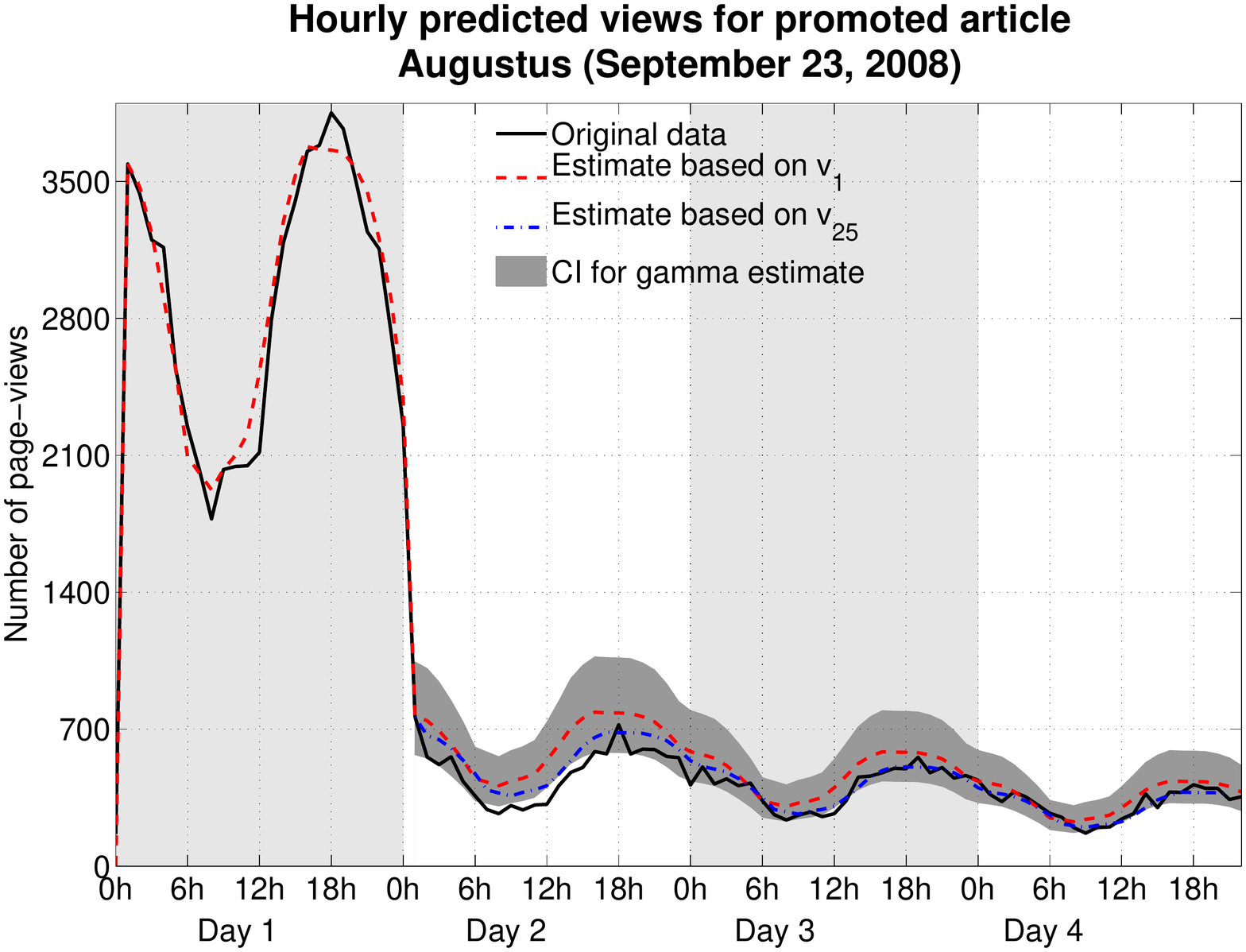}
\includegraphics[width=\columnwidth]{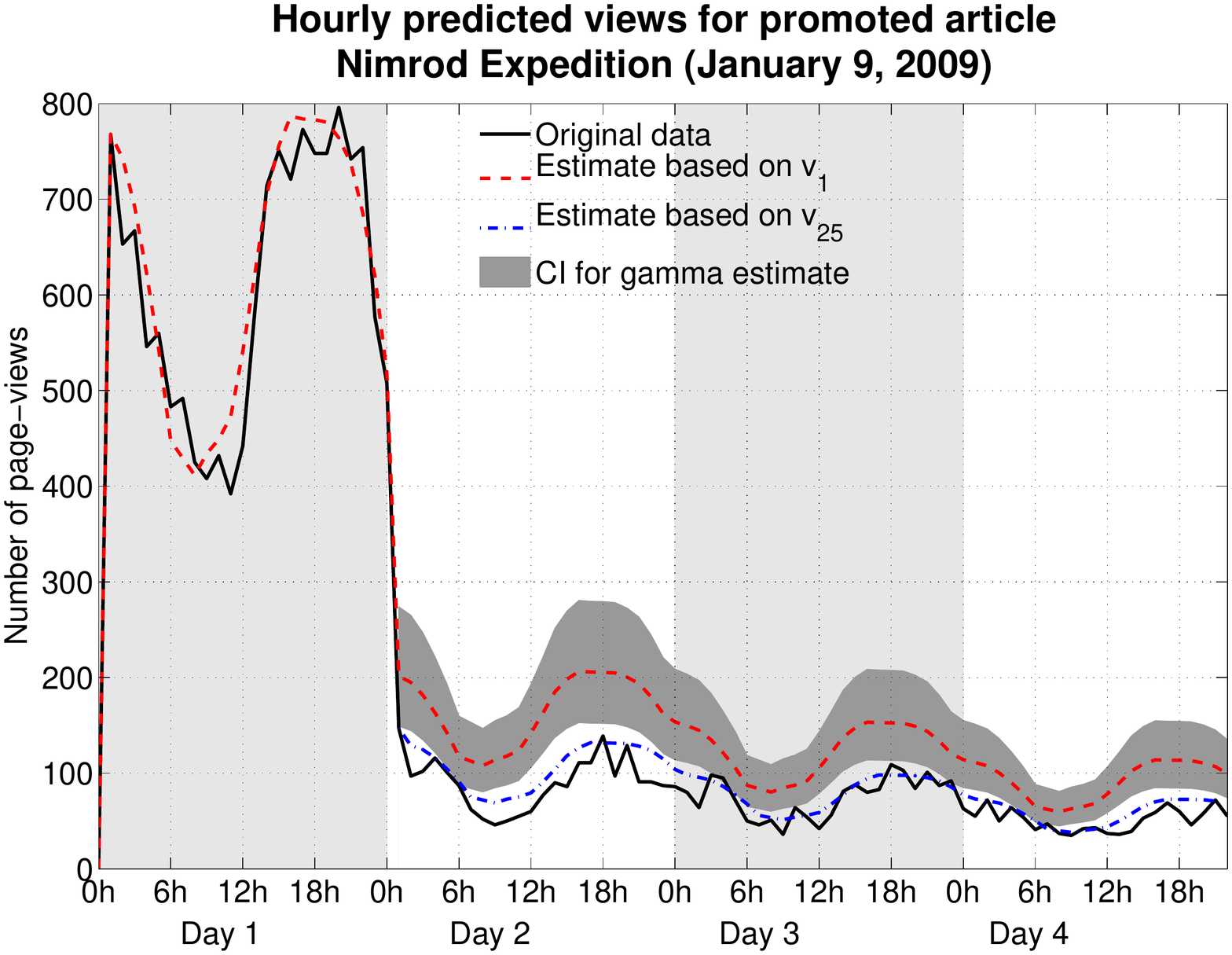}
  \caption{Examples of the prediction of the page-views for a promoted
    article.\label{fig:art_estimate}}
\end{figure}

In Figure~\ref{fig:art_estimate} we present two examples for the
prediction of the popularity for both of the above-defined prediction
methods. We show the initial prediction in red, the interval
$[h(v_1)-\sigma,h(v_1)+\sigma]$ as dark grey area and $\hat{v}_t$
based on $v_1$ and $v_{25}$ in blue. While the prediction of the
article {\it Augustus} performs well already using only $v_1$, similar
prediction overestimates the views of the article {\it Nimrod
  Expedition}.

\begin{figure}[!tb]
\centering
\includegraphics[width=\columnwidth]{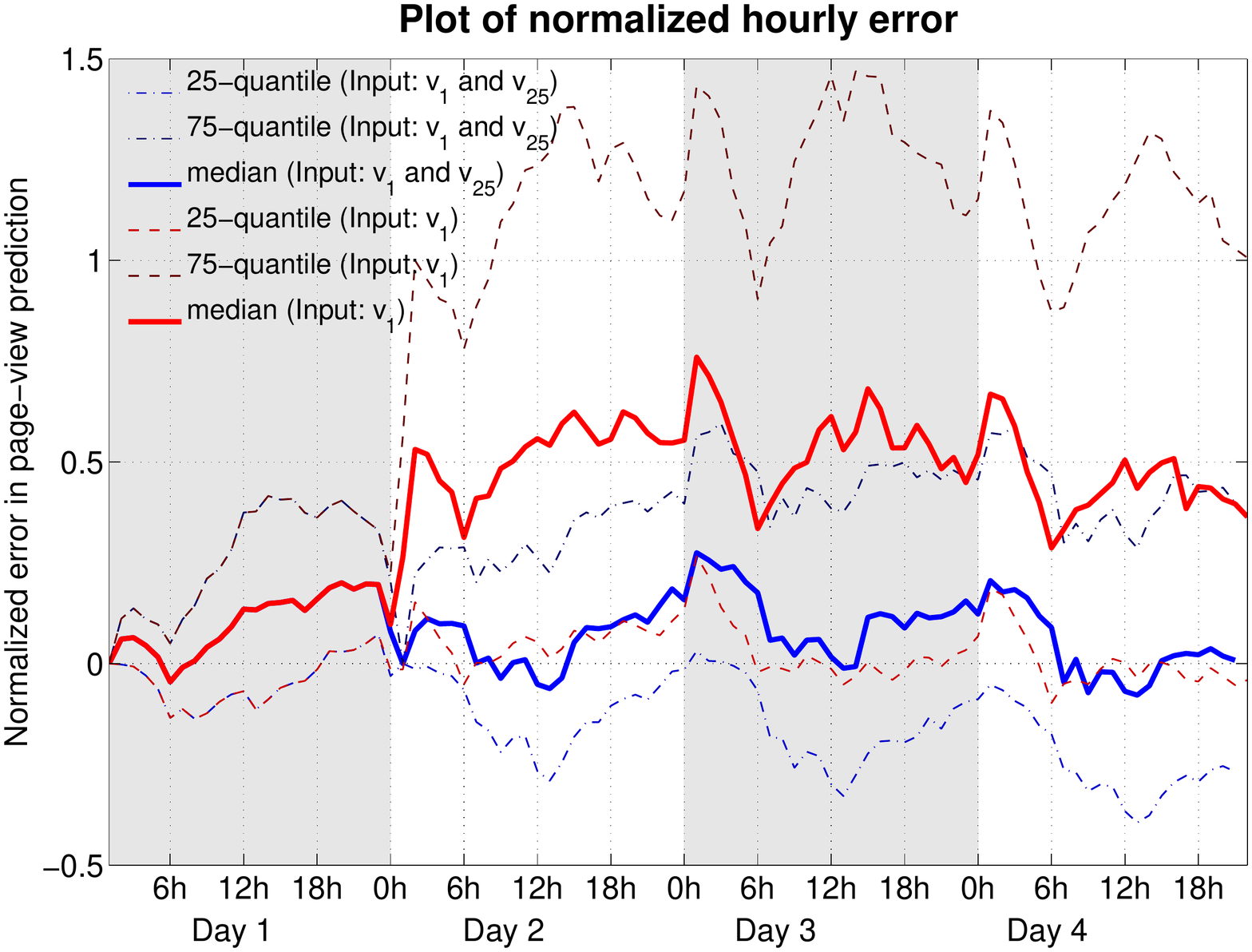}
\caption{Hourly normalised errors.\label{fig:hourly_perc_error}}
\end{figure}

We analyse the normalised hourly errors $(\hat{v}_t-v_t)/v_t$ for all
articles under study for both prediction methods: errors for just
$v_1$ are plotted in red, while errors using both $v_1$ and $v_{25}$
in blue. From Figure~\ref{fig:hourly_perc_error} we observe that our
prediction performs well for the first day of exposure. We recall that
for this time interval we only use $v_1$ for the prediction. For the
second, the third and the fourth days we observe an increase of the
spread of hourly errors. However, this increase is not present for
the second prediction technique.

\begin{figure}[!tb]
  \centering
  \includegraphics[width=\columnwidth]{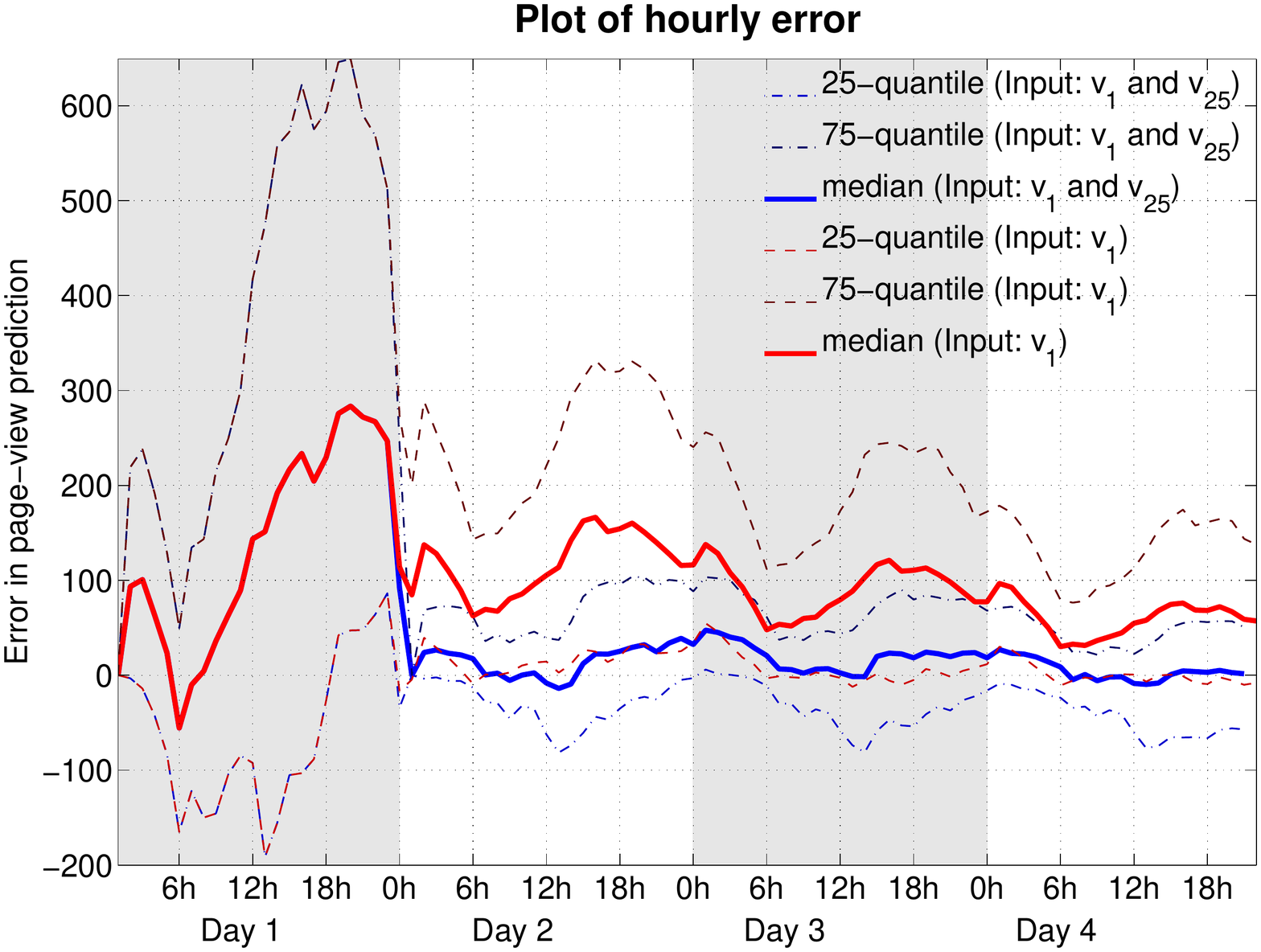}
  \caption{Hourly errors of actual
    page-views.\label{fig:hourly_error}}
\end{figure}

Finally, we present the absolute hourly errors $(\hat{v}_t-v_t)$ in
Figure~\ref{fig:hourly_error}. Interestingly, we observe that the
absolute error towards the second half of day 1 is larger. This is
caused by the fact that we model the negative jump only to occur during one
specific hour whereas for some articles it actually starts a few hours
before the end of the first day of the exposure duration. We also see
that the hourly error during the second, the third and the fourth days
are slightly increasing. 
This is similar to the observation in
Figure~\ref{fig:hourly_perc_error}. Again, the prediction method which
uses both $v_1$ and $v_{25}$ as input outperforms the model that only
uses~$v_1$.

\section{Conclusions and Discussion}\label{sec:conclusion}
We have presented a simple yet powerful model for the view dynamics of
promoted  content on  Wikipedia. The  model shows  that the  number of
views an article receives decays exponentially in time with a constant
decay  rate if  the dependency  of the  data on  Wikipedia's circadian
activity cycle is removed. The  only exception from this decay rule is
the presence  of a  negative jump  when an article  is moved  from the
``today's featured'' to the list of ``Recently featured'' after 24h of
being promoted,  only to decrease  later again with the  same constant
decay rate. The  model allows to predict the  popularity of an article
using only  the number of views  it receives during the  first hour of
exposure. The quality  of the prediction can be  improved if the model
is  updated  right  after  an  article  is  moved  to  the  ``Recently
featured'' section.

Our model, based on the Poisson process, provides a simpler mechanism
to explain page-view behaviour than other recent studies
(e.g. \cite{Wu06112007,Szabo2010}). It should allow to describe and
compare view dynamics on other websites or parts of websites with
similar update strategies, e.g. online newspapers which are updated on
a daily basis, or a list of today's recommended items (mobile apps,
products, etc).

The decay factor can be a useful parameter to account
for the half-live of a piece of content on a given site. The findings
might also be useful to predict the success rate of new online
advertisements or sponsored content in general.

\bibliographystyle{abbrv}
\bibliography{biblio_wikipedia}

\end{document}